\documentclass[aps,prl,onecolumn,showpacs,superscriptaddress,groupedaddress]{revtex4}  
\usepackage{graphicx}  
\usepackage{dcolumn}   
\usepackage{bm}        
\usepackage{amssymb}   

\newcommand{\beq}{\begin{equation}}
\newcommand{\beqa}{\begin{eqnarray}}
\newcommand{\bey}{\begin{eqnarray}}
\newcommand{\eeq}{\end{equation}}
\newcommand{\eey}{\end{eqnarray}}
\newcommand{\eeqa}{\end{eqnarray}}
\def\grad{\nabla}

\def\msun{M_\odot}

\def\lsim{\mathrel{\raise.3ex\hbox{$<$\kern-.75em\lower1ex\hbox{$\sim$}}}}
\def\gsim{\mathrel{\raise.3ex\hbox{$  $\kern-.75em\lower1ex\hbox{$\sim$}}}}

\def\r{{\mathbf r}}
\def\x{{\mathbf x}}

\def\BF{ {} }

\begin{document}

\title{Modified Kepler's Law, Escape Speed and Two-body Problem in MOND-like Theories}
\email[Email address:]{hz4@st-andrew.ac.uk}
\author{HongSheng Zhao}
\affiliation{Scottish University Physics Alliance, University of St Andrews, KY16 9SS, UK}
\affiliation{
UniversitŽ de Strasbourg,
Observatoire Astronomique,
11, rue de l'UniversitŽ,
F-67000 Strasbourg,
FRANCE}
\author{Baojiu Li}
\affiliation{DAMTP, Centre for Mathematical Sciences, University of Cambridge,
Wilberforce Road, Cambridge CB3 0WA, UK}
\affiliation{Kavli Institute of Cosmology Cambridge, Madingley Road, Cambridge
CB3 0HA, UK}
\author{Olivier Bienaym\'e}
\affiliation{UniversitŽ de Strasbourg,
Observatoire Astronomique,
11, rue de l'UniversitŽ,
F-67000 Strasbourg,
FRANCE}
\date{7.7.2010}
\begin{abstract}
We derive a simple analytical expression for the two-body force in a
sub-class of MOND-like theories and make testable 
predictions in the modification to the two-body orbital period,
shape, and precession rate, and escape speed etc. We demonstrate
the applications of the modified Kepler's law in the timing of
satellite orbits around the Milky Way, and checking the feasibility of MOND in the
orbit of Large Magellanic Cloud, the M31 galaxy, and the
merging Bullet Clusters.  MOND appears to be consistent with satellite orbits although with a tight margin.
Our results on two-bodies are also
generalized to restricted three-body, many-body problems, rings
and shells.  
\end{abstract}
\maketitle

\section{Introduction}

The Kepler's law, or the full analytical solution to the two-body
problem is perhaps the most powerful prediction of Newton's
gravitational law, $|{\mathbf F}_1|=|-{\mathbf F}_2|={G m_1 m_2 \over r_{12}^2}$ 
for the forces between any two point masses
$m_1, m_2$ orbiting each other with a separation $r_{12}$. For
bound orbits the Kepler's law predicts the relations between
masses and the orbital period ($T \propto \sqrt{r_{12,max}^3 \over
(m_1+m_2) G}$). 
The total mass of any binary or merging cosmological bodies can be constrained 
by the fact that first a binary must be bound, and 
second the binary's period has to be smaller or comparable to the age of the universe.  
These simple conditions of boundness and timing have powerful applications in 
constraining the unknown physics of the widely-speculated cosmological dark matter.

For example, {\it if} the Milky Way had 
nothing except its baryonic mass $m_1 \sim 5 \times 10^{10}$ solar
masses, then the specific gravitational potential ${G m_1 \over
r}$ would be too shallow to bound the fast moving satellites of
the Milky Way, e.g., the LMC and Sgr with velocity about 300-380
km/s at 50 kpc and 16 kpc from our center well exceeds their local
escape speed of about 100-150km/s. To explain the long tidal tails
around these satellites, one must deepen the potential, e.g., by
adding a dark matter halo, so that these satellites can be on bound orbits 
{\it and} have a few pericenter passages to create the tidal tails.  
There are constraints coming from being bound and constraint
{\BF Another example is 
the colliding Bullet clusters, which shows enormous relative velocity in the shocked X-ray gas  
$\sim 4000$ km/s;  
in comparison if the merging clusters were modeled as two baryonic gas clouds of total mass $\le 10^{14}\msun$ 
falling for the first time from infinity to the present separation of 400-700 kpc, one expects at most  a Keplerian escape speed 
$\sqrt{2G(m_1+m_2) \over r_{12}} \le 1500$km/s.  It has been suggested that even 
the gravity of cosmologically reasonable amount of dark matter is barely enough, and 
an attractive fifth force between dark matter might be necessary to explain the fast speed of the bullet in a Newtonian gravity\cite{FR,SF,LeeK}}.  

Likewise the Kepler's law allows us to time the motions of
celestial objects. For example, in the limit that the local group
can be approximated as two points masses: the Milky Way and M31
galaxy, Kepler's law can be used to estimate their total mass
$m_1 + m_2 \sim (2-3) \times 10^{12}\msun$, one order of magnitude larger than the baryonic 
mass, hence one can argue the existence of dark halos
in these two galaxies under Newton's gravity if $r_{12}$ is 
comparable to the semi-axis of the orbit, and the 
half-period $T/2 \sim 14$ Gyrs so that the two has enough time to
move apart and turn around in a Hubble time.   

{\BF Nevertheless there is another way  to make galaxy potential
deeper without introducing unknown matter}.  
The Modified Newtonian Dynamics (MOND) predicts an
enhancement of the gravitational coupling constant when the
gravity drops below $a_0 \sim 1.2 \times 10^{-10}{\mathrm m/s^2}$,
$\tilde{G}  \sim G  \left[1,{a_0 \over |{\mathbf
g}|}\right]_{\max}$. This produces the effect of a dark halo
without actually invoking real dark matter \cite{BM84}. It is
interesting to derive the modified Kepler's law for the MOND theories,
and contrast their predictions on satellite orbits with Newton's
gravity. Unfortunately, the prediction of the MOND theories generally invokes solving a
modified Poisson equation $\nabla \cdot {{\mathbf g} \over 4\pi
\tilde{G}} =  \sum_i m_i \delta({\mathbf r}-{\mathbf r}_i)$
numerically over all 3-dimensional space for the gravity ${\mathbf
g}$ before obtaining the forces on the two-bodies. The problems
become even harder if there are $N>2$ bodies. To study the orbit,
one would need an N-body code to integrate the orbits step by step
and solve for the modified Poisson equation each step. To date
there have been no realistic simulations of the orbits of two
point-masses in MOND.

Here we show that in some versions of MOND, the virial is a fully
analytical expression for arbitrary matter densities. In the limit
that we can neglect the sizes of the bodies, one can invert the
virial $r_{12} F_{12}$ to find the forces $F_{12}$ between two
bodies.  This completely by-passed the Modified Poisson equation.
The advantage of this analytical result, rigorous in limited versions of the MOND theory, 
is to facilitate the estimation of the modification effects of MOND-like theories.  
E.g., Previous calculations of the timing of the encounter of
the Bullet Cluster in MOND are based 
purely on numerical simulations \cite{AM08,N08,LZK}, which made
the essential physics somewhat more obscure.
Our result allows one to gain similar level of analytical intuition as enjoyed in 
standard Dark Halo theories in Newtonian gravity, albeit the analytics can never substitute 
more rigorous and realistic numerical cosmological calculations and simulations in all these theories.  

It is worth recalling two interesting features in Newton's
gravity: the pericenter of the bodies has no precession because
there is no distinction of a radial oscillation period and a
period to turn 360 degrees in angular direction. The virial of the
two-body system $W \equiv \sum_{i=1}^{2} {\mathbf r}_i\cdot
{\mathbf F}_i$ equals to $ - {G m_1 m_2 \over
r_{12}}$, which applies at any time, whether the orbit is
elliptical or hyperbolic.
{\BF We will discuss how these properties are modified in MOND.
Note that the concept of virial applies in all classical gravity theories, and is more
general than the Newtonian virial theorem, and one can speak of 
the instantaneous virial without any assumption of 
time-averaging or equilibrium state of the objects.  }

The outline of the present work is as follows: in \S2 we give 
the actions for two specific versions theories for MOND, and 
show the simple expression for the virial in these theories.    
In \S3 derives the two-body force from the virial, and give the equation of motion.
\S4 generalizes the results to many-body, rings and shells. 
In \S5 we illustrate the applications of the analytical results in calculating 
the orbits of the Local Group objects.  We summarize in \S6.
 
\section{How to calculate the Virial in Newtonian and MONDian Gravity}

For calculability, we adopt the multi-field version of MOND
according to Bekenstein \cite{BM84,B04} or the recent Qusai-linear MOND
(QMOND) version of Milgrom\cite{M10}. In the low-speed weak field limit,
particles move under 
their Newtonian potential $\Phi_N$ and the MONDian scalar field potential
$\Phi_s$ (which plays the role of Dark Matter potential).  These two potentials  
are given by following Poisson equations for an N-body system 
\bey
\grad^2 \Phi_N = 4 \pi G \rho &,& \grad^2 \Phi_s = \grad \cdot (\nu_{QMOND} \grad \Phi_N),  ~\nu_{QMOND}  = \left( {|\grad \Phi_N| \over a_0}\right)^{-1/2},~{\mathrm in~QMOND} \\
\grad^2 \Phi_N = 4 \pi G \rho &,& \grad \cdot (\mu_{MOND} \grad
\Phi_s) =\grad^2 \Phi_N, ~\mu_{MOND} = \left( {|\grad \Phi_s|
\over a_0}\right),~{\mathrm in~MOND}, \eey where the
$\nu_{QMOND}$-function or the $\mu_{MOND}$-function leads to a
deep-MOND effect, and the matter density, \beq \rho(\r) =
\sum_{i=1}^N  {m_i \over V(\r-\r_i(t),b)} \sim  \sum_{i=1}^N
L_i/c^2 , \eeq consists of N softened particles, each with 
the Lagrangian density 
\beq
L_i \equiv \left[ c^2 + {1 \over 2} \left({d \r_i \over d t} - {\r_i d a \over adt}\right)^2 - \Phi_N(\r_i)- \Phi_s(\r_i) \right]  {m_i \over V(\r-\r_i(t), b)} 
\eeq
where $m_i/V(\r-\r_i(t),b)$ is a spherical density profile with a  softening
radius $b$ of the particle center $\r_i(t)$, e.g.,  a so-called
Hernquist  density profile. \footnote{The Hernquist profile has a
density $m_i/V(\r-\r_i(t),b) = {m_i  b \over 2 \pi r (r+b)^3}.$
 Alternatively one could use
a so-called top-hat density profile, where the  volume factor
$V={4\pi b^3 \over 3}$ within a  softening radius $b$ of the
particle center $\r_i(t)$, and zero outside. For N-body system
with a softening radius $b \sim 1$kpc, and a finite total mass
($\le 10^{15}$ solar masses), we can guarantee a maximum of the
Newtonian gravity $|\grad \Phi_N| \le { G \sum m_i \over b^2} \le
10^5 a_0 $  to hold everywhere. This way we exclude very strong
gravity configurations near star-like point masses, where some
corrections of our lagrangian might apply, e.g., one needs to
suppress the scalar field inside the solar system, the edge of the
solar system has a gravity of $10^5a_0$. Here we have also taken
into account of  the cosmic expansion factor $a(t)$ with the ${da
\over a dt} r_i$ term, and the rest energy density ${m_i c^2 \over
V(\r-\r_i(t), b)}$, all quantities are with respect to proper
coordinates $\r$.}  The softened particles ensure that we do not
over-generalize the above theories to situations of strong
gravity.

The above modified Poisson equations can be obtained
self-consistently from minimizing the following action with
respect to $\Phi_N$ or $\Phi_s$: \bey
S_{QMOND} &=&  \int dt d\r^3 \left\{ \sum_{i=1}^N L_i  - \left[ {|\grad \Phi_N|^2 \over 8 \pi G } + {\grad \Phi_s \cdot \grad \Phi_N \over 4\pi G} + {|\grad \Phi_N|^{3/2} a_0^{1/2} \over 6 \pi G } \right] \right\},\\
S_{MOND}       &=&  \int dt d\r^3 \left\{ \sum_{i=1}^N L_i  - \left[ {|\grad \Phi_N|^2 \over 8 \pi G } + {|\grad \Phi_s|^3 \over 12\pi G a_0} \right] \right\}
\eey

The equation of motion for both versions of MOND is derived by
variation of the action $S$ with $\r_i$, which gives \beq {d \over dt}
{d\r_i \over dt}  + {\r_i d^2 a \over adt^2}   = -{\partial
(\Phi_N+\Phi_s) \over \partial \r_i} \eeq  Here the
particles are coupled to the total potential \beq \Phi \equiv \Phi_N +
\Phi_s \eeq which is consisted of two parts, the Newtonian part
$\Phi_N$ and the scalar field part $\Phi_s$.  The gradient of the scalar field 
$-\nabla \Phi_s$ in both versions of MOND plays the
effective role of the dark matter gravity \cite{WuLMC,ZF10}, which is predicted here 
of the amplitude $(Ga_0\sum_i m_i)^{1/2}/r$ at distances $r$ far from the N
particles, i.e., a test particle at a large distance $\r$ would
accelerate with \beq -{\partial (\Phi_N+\Phi_s) \over \partial \r}
\approx  -\sum_{i=1}^N{Gm_i} {\r \over r^3} -
(\sum_{i=1}^N{Ga_0m_i})^{1 \over 2} {\r\over r^2} \eeq

To see that our formulation indeed recovers the Newtonian and deep-MOND limits in a static universe, 
we note that at small radii from a mass $m_i$, a test particle's acceleration $g$ goes as $Gm_i r^{-2}$, 
and at large radii the acceleration $g$ goes as $(Ga_0m_i)^{1 \over 2} r^{-1}$.  
With a bit of algebra one can show the equivalent 
$\mu$-function in MOND is given by $\mu(g) ={g_N \over g} =\nu^{-1}
=1 - \left[ {1 \over 4}+ \sqrt{ {1 \over 4} + {g \over a_0} } \right]^{-1} $ 
if one can assume the distribution of the matter and gravity has spherical symmetry \cite{ZF2006}.  

\section{Application to two-body problem in general}

Zhao \& Famaey \cite{ZF10} found that, in the absence of cosmic
expansion, the virial $W$ and its Newtonian counterpart $W_N$
satisfy a simple relation \beq
 |W| = \int_0^\infty dr^3 \rho \r \cdot (\grad \Phi_N + \grad \Phi_s) = |W_N| + {2 \over 3} \sqrt{G a_0 (\sum m_i )^3}, ~ |W_N| = \int_0^\infty dr^3 \rho \r \cdot \grad \Phi_N
\eeq
which applies to both QMOND and multi-field MOND for an isolated matter distribution in any geometry.

We now apply this to a general two-body system. Following
Milgrom\cite{M94}, we argue that the total virial can be broken
apart into the internal part and orbital part. In the limit the
bodies are separated with distances much bigger than their sizes so that the mutual gravity
is much weaker than the internal gravity in the vicinity of each
(compact) mass, we can neglect the external field when applying
Poisson's equation inside each body, each body can be treated as
isolated system, hence each satisfies its own virial theorem.

From this we can estimate the internal virial,
\beq
|W_i| = \int dr^3 \rho \r \cdot \grad \Phi_{N,i}   + {2\over 3} \sqrt{G m_i^3a_0}.
\eeq
Subtracting off the internal virial $\sum_i W_i$ from the total virial $W$, we find the interaction virial $r_{12}F_{12}$
is given by
\beq
 {\mathbf r}_{12} \cdot {\mathbf F}_{12}=  {Gm_1 m_2 \over r_{12}}
 + {2 \over 3} \sqrt{G(m_1+m_2)^3a_0} - \sum_{i=1}^2 {2\over 3} \sqrt{Gm_i^3a_0}.
\eeq
where we 
have used the fact that the total Newtonian potential energy
subtracting the Newtonian potential energy of each body yields just
the mutual potential energy.
As far as the particles have very small sizes compared to their separation,
we have $r_{12} \cdot F_{12} =r_{12} F_{12} $, where $r_{12}$ has a negligible spread in distance.
So we find that
the mutual force \beq\label{F12}
F_{12}  =  {Gm_1 m_2  \over r_{12}^2 }  +
{ \Xi   \sqrt{G(m_1+m_2)^3a_0}\over r_{12}}, ~\Xi \equiv { 2
\over 3 } \left( 1- \sum_{i=1}^{2} \left({m_i\over
m_1+m_2}\right)^{3/2} \right) \eeq 
Clearly the force is Newtonian at short distance, and at large distance 
tends to a MONDian 1/r force with a non-trivial normalization.

The accelerations of the bodies ${{\mathbf F}_{12} \over m_1}$ and $-{{\mathbf F}_{12} \over m_2}$, 
or the gravity ${\mathbf F}_{12}$ between bodies are actually {\it independent} of their
relative velocities and history, hence the expression is general
for the mutual (MONDian) gravity between two (widely-separated
compact) bodies.  The above expression for force is rigourous if
the sizes of bodies are smaller than their distance. In reality
the (spherical) particles can have a finite size, say $b$, so that the distance between two bodies have 
a distribution of width $\sim b$ instead of a single value $|{\mathbf r}_{12}|$.
Extended bodies also
introduce new effects, such as studied in \cite{Dai} in the
context of the Birkhoff theorem.  Nevertheless one could adapt the
formulae to approximate the effect of of non-spherical bodies or
softened particles.  Such a formula is given in Appendix 
for two axisymmetrical particles with a bulge of length scale $b$ and disk scale length $k$.
Especially in calculating the force between two spherical particles, 
one replaces $r_{12} \rightarrow r_{12}+b$ in eq.~\ref{F12}.  
In all cases we ensure a construction of MOND two-body force that is {\it conserving
momentum, energy and angular momentum of the whole system}.

Note that the two-body force here is far from obvious.   A naive application of MOND could often lead 
to {\it incorrect} answer, e.g., $F_1={G m_1 m_2 \over r^2} + m_2 \sqrt{G m_1 a_0 \over r^2}$
or $F_2={G m_1 m_2 \over r^2} + m_1 \sqrt{G m_2 a_0 \over r^2}$, which would violate momentum-conservation
\cite{AM08}.   Also our two-body result does not hold rigorous if adopting the Bekenstein-Milgrom (BM84) 
theories of MOND where the scalar field is not used, e.g., the calculations in \cite{Dai}.   
In general the forces must be computed 
numerically by first solving the Poisson equations and then integrating the force over the volume of 
the (extended) body concerned.  Our analytical result here helps to calibrate numerical grid or boundary effects
in a numerical code.  

\subsection{Two-body equation of motion in cosmological background}

We are almost ready to apply our derived force to the two-body
problem except that we have to consider the effect of the Hubble
expansion, which is non-negligible for any timing arguments.
Considering the expansion of the background universe $a(t)$ so
that the relative distance of particles 1 and 2 in proper
coordinates  $\r_1 -\r_2 = (\x_1 -\x_2) a(t)$, we find the
equation of motion is given by
\beq {d \over dt} {d \r_i \over dt} = {d^2a \over a dt^2} \r_i  +
\left({{\mathbf F_i} \over m_i}\right). \eeq Note here the
frictional term ${d a \over dt} {d{\mathbf x}_i \over dt}$ in
equations in co-moving coordinates has canceled itself when the
equation is written for the proper coordinates ${\mathbf r}_i$.
Approximate the scale factor as $a(t) =(t/t_0)^{n}$, then the
cosmological term is $n(n-1) t^{-2}$, and is zero if $n=0$
(static) or $n=1$ (empty open universe). For LCDM parameters, the
Hubble parameter $ {da \over a dt} = {1 \over  14 {\rm Gyr}}
\sqrt{0.667+0.333a^{-3}}$, we find the following approximation,
$a(t) \propto (e^y-1)^{2/3}$, where $y \equiv t/11{\rm Gyr}$ and
${d^2a \over a dt^2} =  {2 \over 3 (11 {\rm Gyrs})^2} e^y ( {2
\over 3} e^y-1) (e^y-1)^{-2} $ so that we expect a nearly constant
${d^2a \over a dt^2} ={4 \over 9 (11 {\rm Gyrs})^2}$ at late times
and in the future, and ${d^2a \over a dt^2}=-{2 \over 9 t^2}$ at
earlier times when $a(t) \propto t^{2/3}$. Generally the
cosmological term is an attractive force in matter-domination, and
repulsive in vaccum domination. For the latter part, we shall
consider only a universe dominated by a pure vaccum energy density
$\rho_c={3 H_0^2 \over 8 \pi G}$, the cosmological term \beq
{\mathbf g}_C  = {d^2a  \over adt^2} \r_{12} = \tau^{-2} \r_{12}, 
\quad \tau  \equiv \left({8 \pi G \rho_c \over 3}\right)^{-1/2}  \eeq acts as a static repulsive potential force
for an exponentially growing scale factor $a(t) = \exp(t\sqrt{8
\pi G \rho_c/3})$ for positive $\rho_c$.  We adopt a vacuum universe with the cosmic e-folding time $\tau \approx 11$ Gyrs.

The two-body relative acceleration
\bey
g_{12} &=&  \tau^{-2} r_{12} - \left[{F_{12} \over m_2} \right] {m_1+m_2 \over m_1}\\ 
             &=& \tau^{-2} r_{12} - {G (m_1+m_2) \over r_{12}^2}  -
{ \sqrt{G a_0}\over r_{12}} { 2 (m_1+m_2) \over 3 (m_1 m_2) } \left( (m_1+m_2)^{3/2} - m_1^{3/2} - m_2^{3/2} \right),
\eey

\section{Many-body problem in MOND}

\begin{figure}
\leftline{
\includegraphics[angle=0,width=5cm]{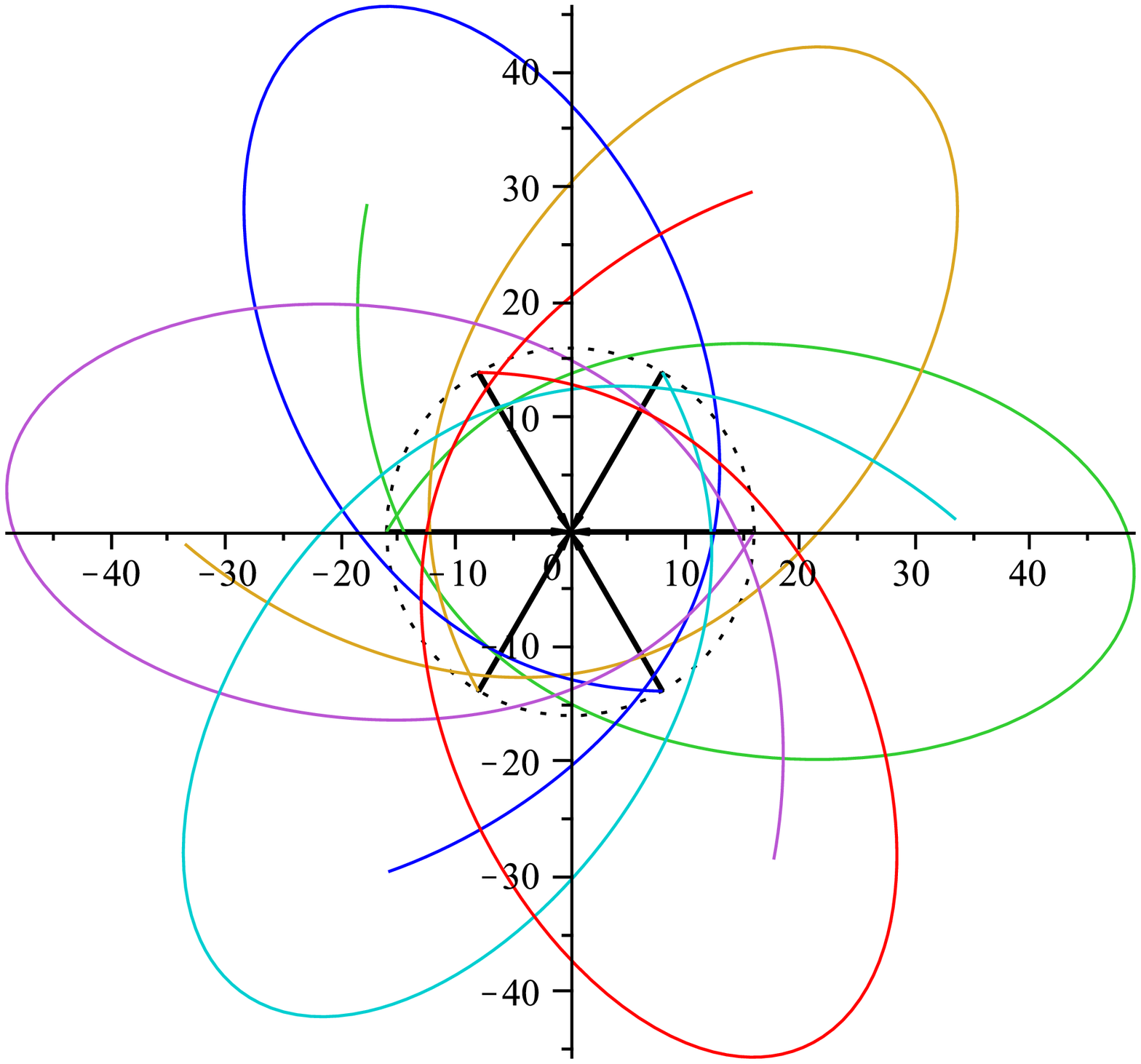}
\includegraphics[angle=0,width=5cm]{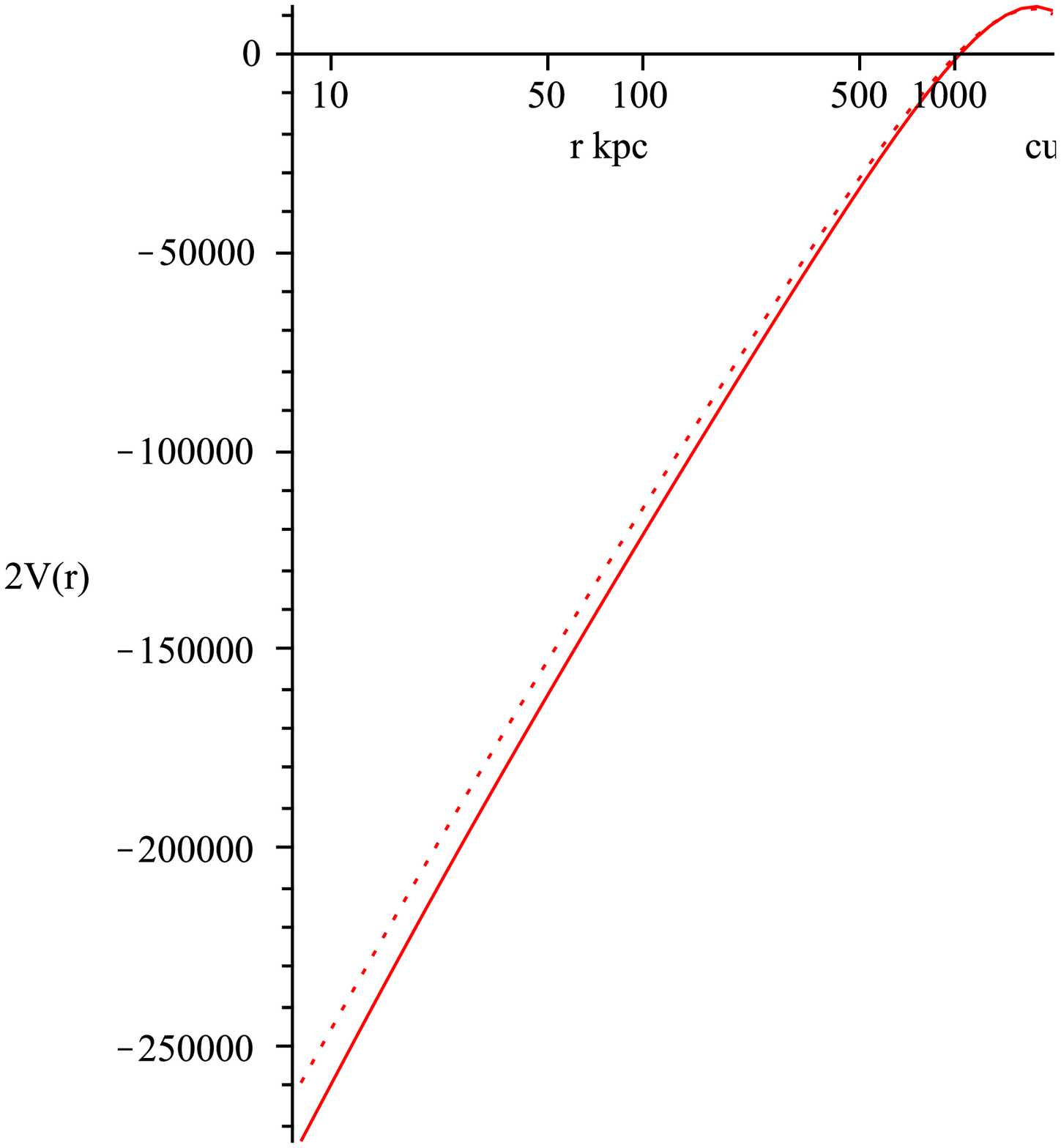} }
\caption{ 
shown is an example of $N=6$ self-gravitating
particles each with mass $m_i = M/(50N)$ initially on a ring of radius 16 kpc moving 280 km/s tangentially   
around the Milky Way (red point at 
origin $M=5\times 10^{10}$ solar masses);  the arrows indicate the force towards the central mass; 
one can see the precession of the pericenter and the stretching of the ring for
the past 1 Gyrs. The right panel shows a semi-log plot of the Milky Way's potential $2V(r)$ (in units of square km/s) 
for a tracer particle (solid) and $N=6$ massive particles on a ring of mass $M/2$ (dashed line) as function of its distance $r$, where we have labeled the location of the cut off radius, where $2V(r_{cut}) \sim 0$;  This allows some stars to escape.  In contrast stars are unable to escape from a logarithmically divergent potential of an isolated MOND galaxy in an empty universe.   All length units are in kpc.}\label{fig:sgr}.
\end{figure}

Consider a 2D or 3D symmetric distribution of $N >1$ identical 
particles of mass $m_i=m/N$ plus a central particle of mass M. As
long as the symmetry guarantees that the forces on the particles $m_i$
is pointing to the central mass M, and that particle $M$ does not 
experience any force, we can use the same virial theorem to obtain 
a relation of the acceleration $g$ of  each particle at $|{\mathbf r}_i|=r$, 
and the Newtonian acceleration $g_N$ and the cosmological
acceleration $g_C$ as follows 
\beq 0 + \sum_{i=1}^{N} {\mathbf r}_i \cdot m_i ({\mathbf g} -{\mathbf g}_N-{\mathbf g}_C)  = {2 \over 3} \left[ (G (M +m)^3 a_0)^{1/2}  -
(G M^3 a_0)^{1/2} -  \sum_{i=1}^{N} (G m_i^3 a_0 )^{1/2} \right]
\eeq Note that the first zero stands for the zero-virial acting on
the central mass M. Also note that any non-symmetry or finite sizes
would prevent us inverting the virial for the force because the
force on each particle would not all be of the same amplitude and
pointing radially.   Such symmetric configurations are generally
contrived with no counterparts in reality, except for perhaps
$N=2$ for binary systems, and $N \sim \infty$ for astronomical
rings and shells, which are fairly common due to mergers of
galaxies, which can form polar rings around spiral galaxies, and
shells around elliptical galaxies.

Making a correction for the expansion of the vacuum only universe of 
a constant cosmic density $\rho_c$, and taking
into account of the finite size $b$, we get \bey \label{eom}
g = {d^2 r \over
dt^2} + {j^2 \over r^3} &=&   \tau^{-2}  r - {G (M +
\tilde{m}_N) \over (r+b)^2}  - {1 \over r+b } (G  \tilde{M}
a_0)^{1/2} \eey where $j$ is the specific angular momentum of the
system, which can be solved from the pericenter radius $r_{peri}$,
where $dr/dt=0$.    Here ${\tilde{M} \over M}$ is a dimensionless parameter depending on the
mass ratios and the detailed geometry of the system and  $\alpha_N \equiv \tilde{m}_N/m$ is a
geometrical factor taking into account of the addition of
Newtonian forces among the $N$ particles. \bey 
\tilde{M}^{1/2} & \equiv &  
{ (M+m)^{5/2} \over Mm} \Xi_1 ~\mbox{if~$N=1$}\\
& \equiv & { (M +m)^{3/2}  \over m} \Xi_N  ~\mbox{otherwise,}
\eey
{\BF where $\Xi_N= {2 \over 3}\left[1 - {  M^{3/2} +  m^{3/2} N^{-1/2} \over (M +m)^{3/2} }\right]$.   
Note that the $N=1$ case is special because the particle M would move around the center of the mass to balance with the single particle $m$ conserve total momentum, 
while the particle M coincides with the center of mass for $N>1$ symmetrically distributed particles of identical mass $m/N$.
 In the appendix we give two expressions which shows the asymptotic behavior more clearly in 
cases with extreme mass ratio.}   We find $\tilde{M} \rightarrow M$
for small mass ratio $m/M \rightarrow 0$, and $\tilde{M}
\rightarrow (1-N^{-1/2})^2 (4/9)m$ for large mass ratio $m/M
\rightarrow \infty$.  If $N=1$, we have  $\tilde{M} \rightarrow M$ or $m$ for extreme
mass ratios.  It can be shown that $0.6 (M+m) < \tilde{M} \le (M+m)$ for any mass ratios and $N$.

For N point masses distributed on regular polygon, we find \bey
{\tilde{m}_N \over m}
&=&1, ~\mbox{if ~$N=1$} \\\nonumber
 &=&  \sum_{i=1}^{N-1} (4N \sin {i \pi \over N})^{-1},~\mbox{otherwise} 
\eey specifically $\alpha_2={1 \over 8}$ for $N=2$,
$\alpha_3={\sqrt{3} \over 9}$ for $N=3$, $\alpha_4={1+ 2\sqrt{2}
\over 16}$ for $N=4$, $\alpha_6={5 \over 24} + {\sqrt{3} \over
18}$ for $N=6$ etc. etc.. By choosing a large $N$ we can
essentially model a circular "ring" of self-gravitating particles.
A similar calculation can be done for $\alpha_N$ if the points are
distributed on regular polyhedra; {\BF the expressions are given Appendix C. } 
For large $N$ the configuration is a fair approximation to a spherical "shell"
distribution of self-gravitating particles.

The above analytical results reveal several interesting
distinctions of the MOND force with Newtonian force. (i) On a ring
or a polygon with $N\rightarrow \infty$, the $G\tilde{m}/r^2$ term
in the Newtonian force diverges, and $\sqrt{Ga_0\tilde{M}}/r$ term
in the MOND part approaches a common constant rotation curve
because the MONDian $\tilde{M}^{1/2}/M^{1/2}$ term depends on the
mass ratio $m/M$ and linearly on $N^{-1/2}$, while in Newtonian
the $\tilde{m}/m$ term depends purely non-linearly on the geometry
parameter $N-1$. (ii) The MONDian force gives the same flat
rotation curve while the Newtonian force produces Keplerian
rotation curves of different amplitudes whether the N particles
form a polygon/ring or a polyhedron/shell. (iii) The MONDian force does
produce precession while the Newtonian force does not.

\subsection{Effective potential, orbital period and precession}

The equation of motion eq.~\ref{eom} can be reduced to purely radial motion  as 
\beq
{d^2 r\over dt^2} = -{d \over dr} V (r), ~
V(r) \equiv  {j^2 \over 2 r^2} - {r^2 \over 2 \tau^2}  - {G (M + \tilde{m}_N) \over r+b}  +  (G \tilde{M}  a_0)^{1/2}   \ln (r+b)
\eeq
where we absorb the centrifugal force into an effective potential $V(r)$,
$j$ is the specific angular momentum of the system. 
We can integrate the equation of motion using energy conservation to get
\beq
{1 \over 2}\left({dr \over dt}\right)^2 + V(r) = V(r_{apo}) = E.
\eeq

For non-radial motion with an angular momentum barrier the
effective potential relates also the pericenter  with the
apocenter via $V(r_{peri})=V(r_{apo})=E$, where $dr/dt=0$.
The time from the pericenter to the apocenter 
and then turn back to the pericenter is given by \beq T_{radial} =
\int_{r_{peri}}^{r_{apo}} {2dr \over |dr/dt| }. \eeq 
Each of the particles on a polygon/ring  will make a rosette orbit, which can precess while keeping the
configuration self-similar;  the whole pattern resembles a
closing/opening shutter.  The backward precession angle per
orbit is determined by\cite{BT} \beq \Delta \phi = 2 \pi -
\int_{r_{peri}}^{r_{apo}}   {j \over r^2} {2dr \over (dr/dt) }
\eeq
In the case that the N particles are on a circular orbit of speed $v_{cir}$,
we have a potential of N-fold rotational symmetry with a
pattern rotation angular speed $\omega_{pattern} =  v_{cir}(r)/r$
and the rotation curve $v_{cir}(r)$ is given by \beq
v_{cir}(r) =  \left[ -{r^2 \over \tau^2}  + {G (M + \tilde{m}_N) \over (r+b)}  +  {r \over r+b} (G
\tilde{M}  a_0)^{1/2} \right]^{1/2} \eeq where the angular
momentum $j$ is found by requiring the effective potential
satisfies $dV/dr=0$ at radius $r=r_{apo}=r_{peri}$. Clearly the
rotation curve $v_{cir}(r)$ is flat at large radii if there is no
cosmological background. The pattern rotation period or pattern
speed of the potential is especially useful for modeling a
self-consistent $m=N=2$ bar potential in MOND.
{\BF Note that the precession here does not change the orbital plane fixed by the constant angular momentum vector, 
only the direction of the pericenters, in some sense 
similar to mercury's pericenter precession due to GR corrections to the Kepler force 
of two bodies.  Any precession of the orbital plane would be the effect of non-spherical shapes 
of the bodies, which we do not go into in this paper.}

Fig.~\ref{fig:sgr} gives an illustration of the many-body problem in MOND, where the orbital parameters  
are inspired by those of the Sgr satellite, whose mass is broken into a self-gravitating tidal debris on a polar plane around 
the Milky Way.   We approximate the Sgr mass by $N=6$ massive particles on a circle of 16 kpc around the central mass M,   
representing the Milky Way.  One can clearly see the precession of the pericenter, the rotation/expansion/shrinking of the initial ring.  

For radial motion with $j=r_{peri}=0$,  the speed that the particles cross the origin is given by $\sqrt{2(V(r_{apo})-V(0))}$.  E.g., the $N$ particles can move radially on a set of self-similiar regular
polyhedron (an approximation to more realistic shells).
For the radial motion, there is a maximum (or an edge) of the
effective potential due to the repulsive force $g_C$ balancing the
Newtonian and MONDian inward force, \beq 0={dV \over dr}|_{j=0}= \left[  \tau^{-2}  r - {G (M + \tilde{m}_N) \over (r+b)^2}  - {1
\over r +b} (G  \tilde{M} a_0)^{1/2} \right]_{r=r_{edge}}. \eeq 
Neglecting
the $r^{-2}$ term due to Newtonian force, and neglecting the
finite size $b$, we get the edge of the potential is at \beq r_{edge} = (G  \tilde{M}
a_0)^{1/4}  \tau. \eeq 
The particles in the MONDian system will not be bound
if the $N$ particles have a specific energy $E$ above
$V(r_{edge})$, in which case the orbits will generally be
hyperbolic-like while keeping the self-similiarity of the
configuration. 

A radial orbit with a large apocenter at $r_{apo}$ would have a period 
\beq
T_{radial} \approx 2\tau \int_{0}^{1} dx \left[  (x^2-1) + 2y^{-2} \ln {1\over x} \right]^{-1/2} , ~y \equiv {r_{apo} \over r_{edge} } 
\eeq
where we have neglected the finite size $b$ and Newtonian potential, 
and rescaled all length with the radius $r_{apo}$ and $r_{edge}$.
Note that an radial orbit which just escapes with
$E=V(r_{edge})$ will take infinite amount of time to reach
$r_{edge}$ because of the linear decline of the radial speed
$dr/dt \propto (r_{edge}-r)$ near the edge.

\subsection{Escape speed, the edge of potential, and the cutoff of the effective dark halo}

Following \cite{WuLMC}, we can define the effective DM mass $M_{EDM} \equiv (g-g_N) r^2/G$.
This is the equivalent Newtonian mass of the DM to generate the same potential as in MOND.
Neglecting the finite size $b$, we find $M_{EDM}$ 
has a positive part linear to $r$ and a negative $r^3$ part, hence the
corresponding effective DM density $\rho_{EDM} \equiv {d M_{EDM} \over 4 \pi r^2 dr}$ has a
positive $1/r^2$ part , and a uniform negative part $\rho_c-3 P_c/c^2=-2\rho_c$ with the point of
zero effective density at \beq r_{cut} ={ r_{edge} \over \sqrt{3}} = (G  \tilde{M}
a_0)^{1/4}   { \tau\over \sqrt{3} } .\eeq 
At this radius the circular speed $v_{cir}(r_{cut})$ and the total dynamical mass
inside can be estimated by $M_{cut} = v_{cir}^2 r_{cut} G^{-1} = M
+ \tilde{m} + {2 \over 3G} \sqrt{G\tilde{M}a_0} r_{cut} $. A
circular orbit bound at $r=r_{cut}$ has  a period $T_{cir}=\sqrt{2}\pi \tau$.  For example the cut off radius of the Milky Way 
is about 2000 kpc, where the peak of the effective potential curve is (see Fig~\ref{fig:sgr}).  

A working definition of an escaping orbit is an orbit which reaches $r_{cut}$, since 
a radial orbit which reaches an apocenter $r_{apo}=r_{cut}$ would have a period $T_{radial} \approx  (1.5-1.7) \tau
\sim 16-19$Gyrs.  Such an orbit takes slightly too long to return over the Hubble time 14Gyrs.   
So the escape speed at any radius $r$ is defined as 
\beq
v_{esc}(r)= \sqrt{2V(r_{cut})-2V(r)}|_{j=0}.
\eeq
E.g., we can estimate the escape speed near the solar neighborhood $r=8{\mathrm kpc}$
adopting $r_{cut} \sim 1500$kpc and  $(G M a_0)^{1/4} =180$ km/s for the Milky Way.
Our analytical formulae predicts a local escape speed 
$v_{esc} \sim \sqrt{2\ln{r_{cut} \over r}} \times (G \tilde{M} a_0)^{1/4}$ km/s $\sim 580 \left({\tilde{M} \over M}\right)^{1/4}
\sim 580 \left[ {\tilde{M} \over M}\right]^{1/4} $ km/s, 
consistent with the observed value $\sim 550$ km/s \cite{WuLMC} for solar neighbourhood stars.     
The so-called external field effect is not as critical here as in \cite{WuLMC}, but our prediction depends on   
the cosmological constant and depends on the satellite-galaxy mass ratio $m/M$ through the function 
${\tilde{M} \over M}$, which is unity for a single star orbiting a galaxy; {\BF  interestingly the escape speed is smaller for a more massive satellite in MOND, however, the difference is fairly small as long as the satellite is less massive than the mass of the host galaxy.} (see Fig.~\ref{fig:sgr}).


\section{Timing the Magellanic Cloud, the Andromeda, the Bullet Cluster}

\begin{figure}
\includegraphics[angle=0,width=6cm]{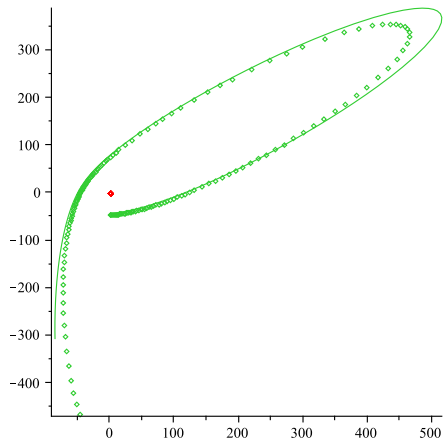} 
\caption{shows the orbits computed for the LMC-MW pair 
with a LMC baryonic mass $M/5$ (small circles) or $M/10$ (lines) respectively for the past
12 Gyrs; the LMC is presently moving to the left with 380km/s at
45kpc below the MW, which is held at the origin (shown by the red
dot).  A generic difference with Newtonian Keplerian orbits is
that the MONDian orbits  depends on the mass ratio, and the direction of the pericenter or apocenter changes 
as the nearly elliptical orbit precess. }\label{fig:lmc}
\end{figure}

\begin{figure}
\leftline{
\includegraphics[angle=0,width=7cm]{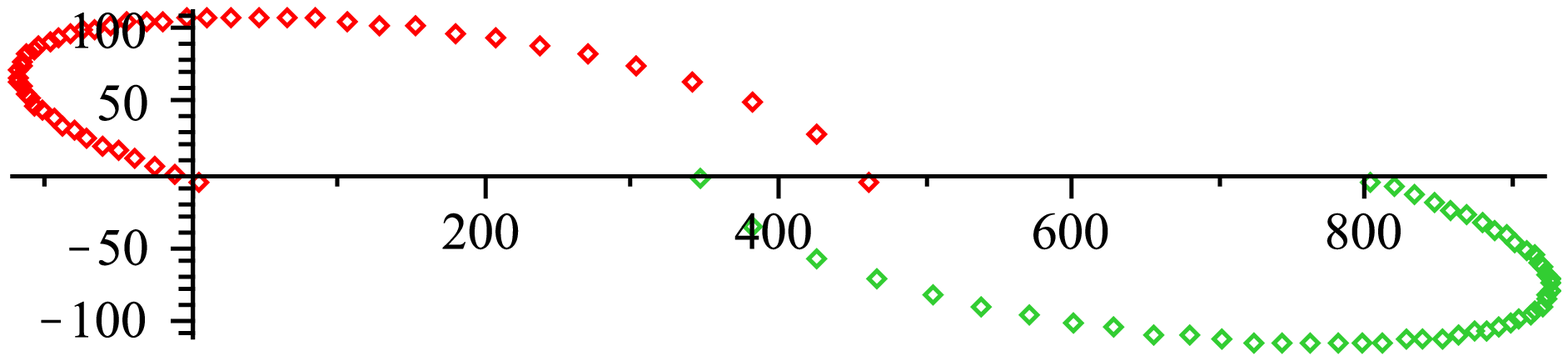} 
\includegraphics[angle=0,width=8cm]{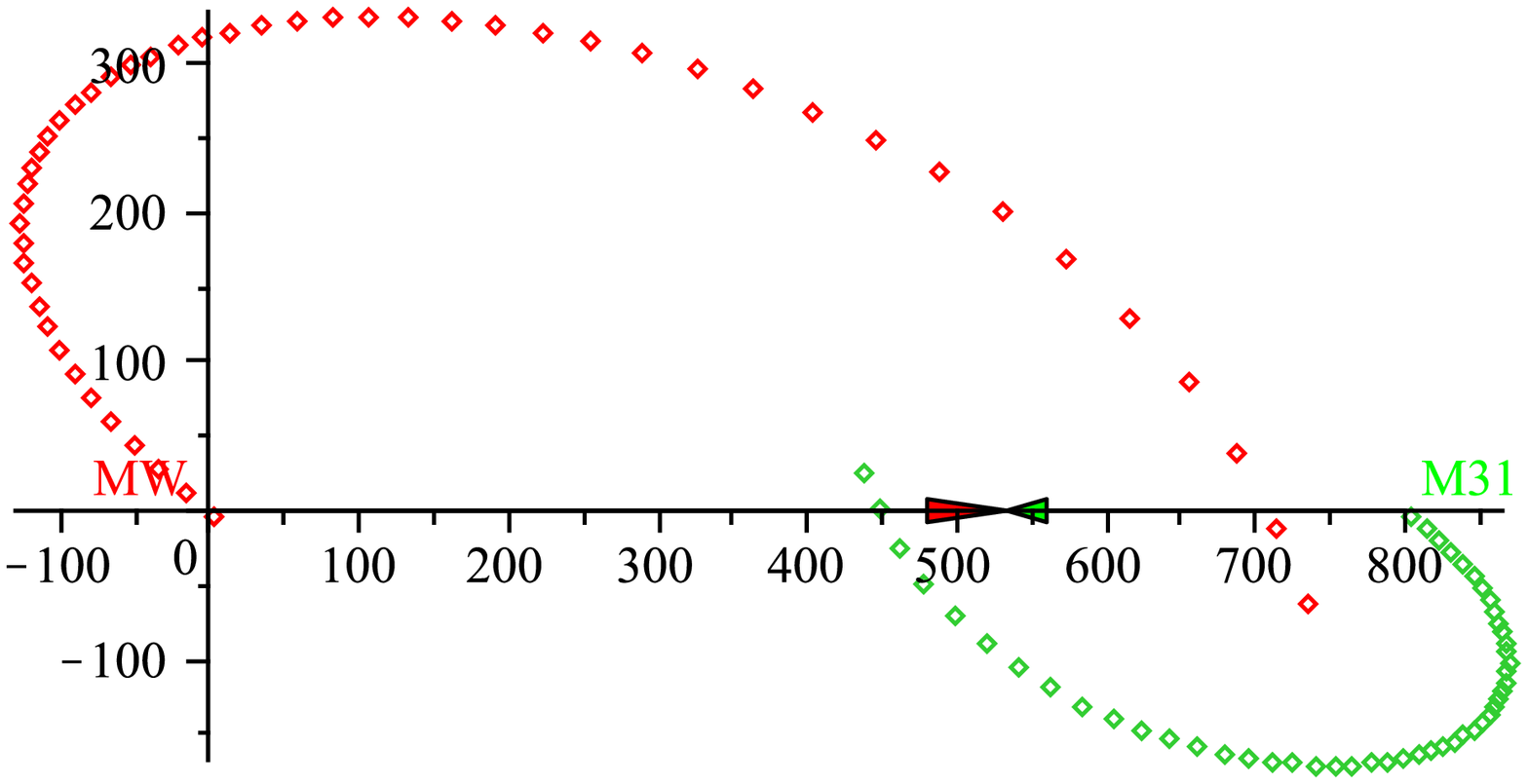} }
\caption{ Shown are two possible orbits for M31 (green)-MW (red)
for the past 12 Gyr: the left panel shows a possible nearly radial orbit with the M31's
baryonic mass the same as that of the Milky Way ($M=5\times
10^{10}$ solar masses) and the right panel shows a significantly non-radial orbit of the
M31 with a mass $2M$. The origin is set at the present position of
the MW, while the M31 is presently 800 kpc away on the right.  The arrows indicate the force towards the center of mass.
}\label{fig:m31}.
\end{figure}

\begin{figure}
\includegraphics[angle=0,width=6cm]{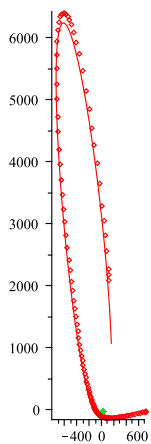} 
\caption{shows the orbit of the Bullet subcluster with respect to
the main cluster (green dot fixed at origin) with a mass ratio 1:2
(red circles) and 1:3 (red line) respectively for 9 Gyrs earlier
than its present age; the Bullet is at 700kpc to the right of the
main cluster and is moving to the right at 4000km/s; both models
assume the same combined mass $M_1+M_2=6 \times 10^{14}$ solar
masses. }\label{fig:bul}
\end{figure}

Timing is a classical argument which assumes two presently neighbouring 
bodies were very close at birth within an expanding Hubble flow away from each other. 
The mutual gravity eventually overcomes the Hubble flow and brings their orbits close to each other again.
Their orbital or dynamical age must then be close to the Hubble time.
A few timing applications of our analytical result is shown in Fig.~\ref{fig:lmc}, Fig.~\ref{fig:m31} and Fig.~\ref{fig:bul}.
The relative distances of these are such that the LMC moves inside the Milky Way, and is presently near the 
pericentre, and the M31 is approaching us from about 800 kpc  
after falling out of the Hubble expansion.

First consider the dynamical age of the Large Magellanic Cloud (LMC), a satellite on an non-circular orbit
around the galaxy.  Recent
observations found the LMC moves with a speed of 360
km/s almost tangentially at a distance 45-50 kpc from the Milky
Way, which is too fast for a Newtonian Dark Halo to keep it
bound\cite{WuLMC}.  In our scenario the MONDian Milky Way has a
baryonic mass $M_1=5 \times 10^{10}$ solar masses, 
and the LMC has a baryonic mass $M_2=M_1/5$, hence $(G a_0 \tilde{M})^{1/4}=180$km/s, we
find $r_{cut}=1500$ kpc, $r_{max}=\exp((360/180)^2/2) \times
45{\rm kpc} = 350{\rm kpc}=r_{cut}/5$.   We estimate the
oscillation period in the radial direction $T_{radial} = 0.3\tau
=4$~Gyrs.  The period in the circular direction is about $1.4$
times longer. This implies that the Large Magellanic Cloud could
be bound and have enough time to circulate the Milky Way more than
once in MOND (cf. Fig.~\ref{fig:lmc}). This is consistent with the standard scenario where
the Magellanic stream is pulled out from the Magellanic Clouds
during one of its pericentric passages of the Milky Way, and possibly interactions among themselves
\cite{LL82}.  {\BF We also note that the baryonic mass of the Milky Way 
adopted for modeling the LMC is roughly consistent with the local escape speed and 
the orbit of the Sgr dwarf.   As shown in Fig.~\ref{fig:sgr},
in particular  our model orbit has a pericenter and apocenter of 10 and 50 kpc respectively, and   
a present 280 km/s tangential velocity at a 
radius 16 kpc from the Milky Way.  These parameters are consistent with the observational data and typical orbits 
found in previous dark halo models \cite{LawM}.  }

We have also computed the orbit of the M31,
whose parameters are not well-determined observationally. The M31 galaxy is
presently at $\sim 800$kpc and is moving at 130km/s towards us.  Its tangential velocity is unknown, 
although perhaps small.  {\BF We study two models}: in
one model M31 has an equal baryonic mass as the Milky Way $M_2=M_1=5 \times
10^{10}$ solar masses and M31 has a nominal small
transverse velocity of 40km/s. In another model M31 is twice as
heavy as the Milky Way\cite{CS} and has a larger transverse
velocity 100km/s to avoid collision with the M33\cite{Loeb}.


Shown in  Fig.~\ref{fig:m31} are the orbits in the past 12 Gyrs, earlier than which the
disks of the Milky Way and M31 are likely not yet formed. MOND
prefers either a small baryonic mass for the Local Group, or a
significant transverse velocity of the M31, something that can be
falsified by future measurements of the transverse velocity. A
transverse velocity helps to keep the pericenter distance large, about 
200-300kpc, hence any tidal effect from the M31 on the LMC is
small, which is perhaps a desirable feature to bound the LMC within 45-450 kpc of
the Milky Way. However, if we adopted a  radial orbit and a large
mass for the M31 as in the standard interpretation of Local Group
timing with a Keplerian orbit of M31 and MW dark halos \cite{BT,
KW}, there would be some tension between the long age of the universe
and short orbital period of M31-MW binary predicted in MOND,
implying that the two systems had an earlier flyby, and are coming
close to each other for the second time.  
Our model cannot yet include any 
acceleration of the whole M31-MW binary towards the Virgo
cluster, which creates 
the so-called external field effect, which can generally reduce 
the MOND force or potential, and lengthen the period
\cite{FBZ,WuLMC}.

To time the Bullet Cluster, which is at $z=0.3$ when the age of
the universe is 10 Gyrs, we can set $T_{radial} \sim 10$
Gyrs. 
A possible orbit is shown in Fig.~\ref{fig:bul}. 
Allowing for some hydrodynamical effect \cite{SF},
we set the speed of encounter $\sqrt{2 \ln{r_{apo} \over r}} (G
\tilde{M} a_0)^{1/4}  \sim (3000-4000)$ km/s at the present
separation $r=700$ kpc, we find $(G \tilde{M} a_0)^{1/4} \ge
3000/\sqrt{2 \ln(7)} \sim 1500$ km/s, hence the matter $\tilde{M}
\ge 2.5 \times 10^{14}$ solar masses, much larger than the combined 
baryonic content $\le 10^{14}$ solar mass
for both systems.  This implies the need for non-baryonic matter, perhaps (sterile) neutrinos in
both systems\cite{Sanders03}. These crude estimates are consistent
with the previous findings \cite{AM08,N08}.

\section{Summary}

In short, previous tests of MOND have largely been limited to
fitting rotation curves or velocity dispersion curves inside axisymmetric
galaxies or galaxy clusters. A few studies in the literature on
non-axisymmetric configurations have relied primarily on numerical
codes \cite{WuLMC,LZK,Dai,Ciotti,Combes}.  The numerical complexity severely limits
our theoretical intuitions on this class of non-linear theory of gravity.

Here we derive {\it the modified Kepler's law analytically} for a
two-body system and for restricted many-body problem in the context of
two versions of theories for MOND. We demonstrate the powerful use
of modified Kepler's law by applying our analytical results to
make predictions of the orbital motions for real systems. These
analytical results are also useful for testing numerical codes
(e.g., \cite{Dai}) and for getting intuitions. It appears that in
the case of the Bullet Cluster a pure MOND without non-baryonic
matter (e.g., neutrinos) is not enough to explain the fast motions
of the bullet. On the other hand the timing of the M31's orbit
would imply an uncomfortably low mass for the M31 unless its orbit
towards the Milky Way has significant amount of angular momentum,
{\BF which is a strong prediction for MOND to survive in the context of the two-body
problem in the Local Group.  
The baryonic mass of the Milky Way 
$\sim 5\times 10^{10}$ solar masses seems enough to consistently explain
the rotation curve \cite{ZF2006}, the local escape speed, and 
the morphology and kinematics of the Sgr stream (Fig.~\ref{fig:sgr}); note in our approximation 
we have neglected the effects of the detailed form the MOND $\mu$ function in these regimes.
The orbit of the LMC is bound around the Milky Way in MOND, however, with only two
pericentric passages in the past 12 Gyrs, so it remains to be seen
if these pericentric passages are enough to generate the detailed morphology of the
Magellanic Stream.  }

{\BF To conclude, our analytical formulae for the MOND gravity in two-bodies 
provide some tools for studying MOND beyond
rotation curve fitting, where MOND has been very successful.  We propose to 
use the timing argument in two-body problem to probe the dynamics of the LMC, M31 and the Bullet clusters 
in MOND.  The ultimate falsification or proof rests on more detailed numerical
modeling with improved kinematic data.}

\appendix
\section{Appendix A: Motion-independent force and finite-size correction}
 
Assume the body $m_1$ is a Kuzmin-Hernquist disk-bulge flattened system  
(as introduced in Shan et al. 2008)\cite{Shan} with two imaginary centers at $|\pm {\mathbf k}|$ above or
below the plane of the disk \cite{BT}, one can apply the
formulae as if  $m_1$ is a spherical Hernquist body of scale length $b$ centered on a point
below the plane whenever the body $m_2$ is above the plane, and
vice versa.  So the equation of motion for $m_1$ and $m_2$ are
\beq
 {m_2 d^2\r_2 \over dt^2}  = - {m_1 d^2\r_1 \over dt^2} ={\mathbf F} =
+ {\partial \over \partial \r_2} {G m_1 m_2\over |\r_2 - \r_1 \pm  {\mathbf k}|+b}
- { 2 \sqrt{G a_0 (m_1+m_2)^3} \over 3 } \left(1- {m_1^{3 \over 2} + m_2^{3 \over 2} \over  (m_1+m_2)^{3 \over 2} } \right)  {\partial \over \partial \r_2} { \ln |\r_2 - \r_1 \pm  {\mathbf k}| +b }.
\eeq

For two spherical particles with $k=0$,  the mutual force \beq
F_{12} =m_1 a_1 = m_2 a_2 =  {Gm_1 m_2  \over (r_{12} + b)^2 }  +
{ \Xi   \sqrt{G(m_1+m_2)^3a_0}\over r_{12}+b}, ~\Xi \equiv { 2
\over 3 } \left( 1- \sum_{i=1}^{2} \left({m_i\over
m_1+m_2}\right)^{3/2} \right), \eeq where we have opted to smooth
our point-like particle with a common scale $b$ by a
Hernquist-smoothing Kernel, which is not always rigorous, but allows a 
non-divergent estimation of forces in situations where the 
bodies are overlapping.  The essential thing is to keep the forces
{\it rigourous for point mass} and that $F_{12}=F_{21}$ in general
such that the system conserves total momentum. The forces on the two bodies are in 
opposite directions, keeping their center of mass
fixed. 

\section{Appendix B: Alternative expression for $\tilde{M}$ in binary configuration, and in symmetric many-body configurations}
Following expressions for the effective MONDian mass $\tilde{M}$ 
have better asymptotic behavior for $N=1$ binary configuration with two masses M and m, 
and for a central mass $M$ plus a $N>1$  symmetric identical particles of mass $m/N$.
\bey \tilde{M}^{1/2}
&\equiv&  (M+m)^{1/2} (1 - {2 M^{1/2}m^{1/2} \over 3 (M+m)}) \left[
{1 \over 2} + {M^{3/2} +  m^{3/2} \over 2(M+m)^{3/2}} \right]^{-1}~\mbox{if ~$N=1$} \\ \nonumber
& = & \left[ { 2 M (M+m)  + (2/3)m^2  \over  (M +m)^{3/2}  + M^{3/2} }
-  {2m^{1/2} \over  3N^{1/2}} \right], ~\mbox{otherwise} \eey  
For equal mass binary $m=M$, and $N=1$, we have $\tilde{M} = 2
\left[ 1- { 2 \over  6}\right]^{2}\left[{1 \over 2}  + {1 \over
2^{3/2}} \right]^{-2} m \sim 1.2 m$. 

\section{Appendix C: expressions for $\tilde{m}_N$ for $N>1$ identical masses distributed on a regular polyhedra}
It is somewhat lengthy but straightforward to calculate the Newtonian force vectors between N particles 
distributed on regular polygons and polyhedra of radius r.  Summing up the forces on each particle, one can find 
the forces are indeed centripedal, the acceleration is given by $G\tilde{m}_N/r$,
The expressions for $\alpha_N ={\tilde{m}_N \over m}$ are found as follows: 
$\alpha_6={1 \over
24}+{\sqrt{2} \over 6}$ for a $N=6$ points on octahedron and
$\alpha_8={1 + \sqrt{3/2}+ \sqrt{3} \over 32}$ for a $N=8$ points
on a cube. There are totally 5 possible regular polyhedra or
platonic solids, with $N=4,6,8,12,20$ vertices for tetrahedron, octahedron, cube,
icosahedron, and dodecahedron.

\end{document}